\documentclass[preprint,aps,amssymb,showpacs,superscriptaddress,nofootinbib]{revtex4}
\usepackage{graphicx}

\usepackage{hyperref} %
\hypersetup{ colorlinks=true, linkcolor=blue, citecolor=red,
filecolor=magenta, urlcolor=cyan }

\newcommand{\Case}[2]{{\textstyle \frac{#1}{#2}}}

\begin{document}

\hfill\today


\title{Linearized gravitational waves in de Sitter space-time}

\author{Ghanashyam Date} \email{ghanashyamdate@gmail.com}

\affiliation{Chennai Mathematical Institute,\\ H1, SIPCOT IT Park,
Siruseri, Kelambakkam 603 103 India}


\begin{abstract}
This is a brief note on a recently flagged issue \cite{CHK} regarding
the precise characterization of quadrupolar truncation of the linearized
gravitational waves in de Sitter space-time.  An apparent inconsistency
was noted while transforming the linearized solution truncated to
quadrupolar contribution to the Bondi-Sachs form. A consistent
truncation was identified as one where the transformation of the
truncated solution does not generate logarithmic terms inadmissible in
the Bondi-Sachs asymptotic form. 

I point out the role of the gauge conditions and the solution of the
homogeneous equation in obviating the apparent inconsistency issue. 
\end{abstract}


\maketitle


%

It is a pleasure and honour to participate in the felicitations on the
occasion of the 80th birthday of Professor Naresh Dadhich. He has been a
continuous source of gentle encouragement, an inspiration and a warm
friendship for close to 3 decades. I wish that he continues spreading
his warmth for years to come.
\section{Introduction} 
\footnote{This is an expanded version of the talk presented at the
National Conference on Classical and Quantum Gravity held at the Cochin
University of Science and Technology.}
After discovering that linearized Einstein equations admit waves, it
provided a new mechanism for gravitationally bound systems to radiate
and loose energy, momentum, angular momentum etc. The absence of
coordinate invariant local definitions to track the flow of these
quantities by gravitational waves and the non-linear nature of Einstein
equation make it very confusing to formulate the notion of radiative
losses.  The problem boils down to defining a consistent set of
asymptotics i.e.  characterization of an asymptotic region, asymptotic
form of appropriate solutions, distinguishing radiation conveying the
energy from the source region to the asymptotic region and of course a
coordinate invariant quantification of the energy etc. For ``isolated
systems'' or compact sources of gravitational fields, the problem of
asymptotics of the Einstein equation with $\Lambda = 0$ was solved
satisfactorily by the works of Bondi, Pirani, Robinson, van der Burg,
Metzner, Sachs, Penrose \ldots .  In essence, gravitational radiation is
to be {\em defined at} conformal infinity, ${\cal I}$, by intrinsic
structure available on it. In suitably defined coordinates in the
vicinity of this boundary the metric components have a specified
asymptotic fall off.  With these asymptotic forms the energy, momentum,
angular momentum and their fluxes are defined in a gauge invariant
manner with appropriate loss formulae. This step is {\em essential} to
be able to compute the energy loss and subsequent shrinking of the bound
source leading to collapse, merger of binaries etc. This in turn is
necessary to find the waveform of gravitational waves emitted during
this process which is what is detected and analyzed in the
interferometers. 

The characterization at infinity still needs to be related to specific
sources. Furthermore, the analysis at infinity uses exact Einstein
equations. The fields near the source region are however computed in a
perturbative manner in the post-Minkowskian (PM) expansion, over and
above the post-Newtonian (PN) series. These solutions are matched with
the asymptotic fields to relate the energy loss to source motion. This
is typically done writing both solutions in the form of a multipole
expansion and matching is done for each order of multipole.  Since the
energy loss is determined in terms of the asymptotic fields which feeds
into the source motion described using PN expansion, consistency of any
truncation up to specific multipole and PN order needs to be ensured.
This has been streamlined for the $\Lambda = 0$ case.  For the de Sitter
background, these steps of going beyond linear order (``post-de
Sitter'') and matching of source region solutions with asymptotic
solutions are not developed yet.

The recent work of Comp\`ere, Hoque and Kutluk \cite{CHK}, for $\Lambda
> 0$, points to a potential problem at the linear level itself.  It
arises in comparing linearized solutions in two different charts of the
Poincare patch of the background de Sitter space-time: the conformal
coordinate $(\eta, \vec{x})$ and the Bondi-Sachs coordinates $(u, r,
x^A$) with $x^A$ denoting the usual spherical polar angles $\theta,
\phi$.  To articulate their observation, as well as to clarify the
computations, it is useful to trace through the usual steps followed in
the conformal chart (or the cosmological chart with the conformal time
$\eta$ replaced by the synchronous time $t$). The transformation to the
Bondi-Sachs coordinates is done in the following section
\ref{BSChrt-sec}. 
\section{Usual computation for $\Lambda > 0$ in Conformal chart}
\label{ConfChrt-sec}
The relevant part of the de Sitter background space-time is the Poincare
patch which is conveniently described in the conformal chart. In this
chart, the metric takes the form, $\overline{ds}^2 =
\Case{1}{H^2\eta^2}\big(-d\eta^2 + \vec{dx}\cdot\vec{dx}\big), \ H^2 :=
\Case{\Lambda}{3}$.  Denoting the de Sitter metric as
$\bar{g}_{\mu\nu}(\eta,\vec{x})$, the linearized metric is written as
$g_{\mu\nu} = \bar{g}_{\mu\nu} + \epsilon h_{\mu\nu}$. In terms of the
trace reversed $\tilde{h}_{\mu\nu} := h_{\mu\nu} -
\Case{h}{2}\bar{g}_{\mu\nu}, \ h := h_{\mu\nu}\bar{g}^{\mu\nu}$ and
$B_{\mu} := \bar{\nabla}_{\alpha} \tilde{h}^{\alpha}_{~\mu}$, the
linearized Einstein equation becomes \cite{DH-1},
\begin{equation}
	\frac{1}{2}\Big[-\bar{\Box}\tilde{h}_{\mu\nu} +
	\big\{\bar{\nabla}_{\mu}B_{\nu} + \bar{\nabla}_{\nu}B_{\mu}
-\bar{g}_{\mu\nu}\big(\bar{\nabla}^{\alpha}B_{\alpha}\big)\big\}\Big] +
H^2\Big[\tilde{h}_{\mu\nu} - \tilde{h}\bar{g}_{\mu\nu}\Big] = 8\pi
T_{\mu\nu} 
\end{equation}
The divergence of the left hand side is identically zero which implies
the conservation of the stress tensor as expected. The equation is
invariant under the gauge transformations:
$\delta_{\xi}\tilde{h}_{\mu\nu} = \bar{\nabla}_{\mu}\xi_{\nu} +
\bar{\nabla}_{\nu}\xi_{\mu} - \bar{g}_{\mu\nu}
\bar{\nabla}_{\alpha}\xi^{\alpha}$.

The equation is reduced to a simpler wave equation by choosing a gauge
i.e. requiring that $\tilde{h}_{\mu\nu}$ satisfy some additional
conditions. Typically, one sets $B_{\mu} = 0$.  We refer to this choice
as the {\em harmonic gauge}. In the resultant wave equation, different
components of $\tilde{h}_{\mu\nu}$ are coupled. A different choice,
namely, $B_{\mu} = 2H^2\eta\tilde{h}_{0\mu}$, referred to as the {\em
generalized harmonic gauge}, leads to decoupled wave equations which are
conveniently solved. In terms of the variables, $\chi_{\mu\nu} :=
H^2\eta^2\tilde{h}_{\mu\nu}$, the wave equations take the form
\cite{VRS, CHK},
\begin{equation}\label{WaveEqns}
	\Box\Case{\hat{\chi}}{\eta} =
	-\Case{16\pi\hat{T}}{\eta}\hspace{0.5cm},\hspace{0.5cm}
	\Box\Case{\chi_{0i}}{\eta} =
	-\Case{16\pi{T_{0i}}}{\eta}\hspace{0.5cm},\hspace{0.5cm} (\Box +
	\Case{2}{\eta^2})\Case{\chi_{ij}}{\eta} =
	-\Case{16\pi{T_{ij}}}{\eta}~.
\end{equation}
Here $\Box = -\partial^2_{\eta} + \partial^2_{i}$ is the d'Alembertian
in Minkowski space-time and $\hat{\chi} := \chi_{00} + \chi_i^{~i}$. The
solutions of these decoupled wave equations {\em have to} satisfy the
generalized gauge condition,
\begin{equation}\label{GaugeCondn}
	\partial^{\alpha}\chi_{\alpha\mu} +
	\frac{1}{\eta}\big(2\chi_{0\mu} +
	\delta^0_{\mu}\chi_{\alpha}^{~\alpha}\big) ~ = ~ 0 \ .
\end{equation}
Some further gauge freedom is still left as both these equations are
invariant under some residual gauge transformations \cite{DH-1}.

The wave equations being inhomogeneous equations, their general solution
is the sum of the particular solution and a solution of the homogeneous
equation: $\chi_{\mu\nu} := \bar{\chi}_{\mu\nu} +
\underline{\chi}_{\mu\nu}$. The particular solution is determined by the
source and has the schematic form, $\bar{\chi}_{\mu\nu}(x) \sim \int
d^4x' G_{ret}(x,x')T_{\mu\nu}(x')$. The source being spatially compact,
the integral ranges over the source region while the solution is desired
for $x$ in the far exterior region. The gauge conditions
(\ref{GaugeCondn}) are required to be satisfied by the {\em combined
solution} $\chi_{\mu\nu}$.

It is a result stated in \cite{VRS} that in the {\em exterior (source
free) region}, the residual gauge invariance can be exhausted by setting
$\chi_{0i} = 0 = \hat{\chi}$. The gauge condition in equation
(\ref{GaugeCondn}), then implies that we can arrange to have $\chi_{00}
= 0$ as well. With these, we have $\chi_{0\mu} = 0$ which is referred to
as the {\em synchronous gauge condition}. Together with this condition,
the generalized harmonic gauge condition reduces to:
$\partial^j\chi_{ji} = 0 = \chi_i^{~i}$ which is a condition on the
spatial components alone. We refer to it as {\em spatial transverse,
traceless} condition \cite{DH-1}. With this complete gauge fixing, it
suffices to focus on the solution of the inhomogeneous wave equation for
spatial components alone together with the spatial TT gauge condition.
{\em We do not need to consider the particular solutions for the
components $\hat{\chi}, \chi_{0i}$}.

The spatial components, $\chi_{ij}(\eta, \vec{x})$ and $T_{ij}(\eta',
\vec{x}')$ are symmetric tensors on constant-$\eta$ hypersurfaces. Such
tensor fields have the general decomposition of the form:
\begin{equation}
	X_{ij}(\vec{x}) = \big[C(\vec{x})\delta_{ij}\big] +
	\big[\big(\partial_i\partial_j -
	\Case{1}{3}\delta_{ij}\nabla^2\big)E\big] + \big[\partial_iV_j +
	\partial_j V_i\big] + X_{ij}^{TT}(\vec{x}) ~ , ~ \partial^jV_j =
	0 \ .
\end{equation}
The $C$ scalar function is determined by the trace of $X_{ij}$ while the
gradient of the scalar function $E$ together with the divergence-free
$V_j$ constitute a vector field determined by the divergence of
$X_{ij}$. The left over $X_{ij}^{TT}$ gives the two independent,
transverse, traceless parts ($TT$ for short) of $X_{ij}$. We apply this
to the particular solution of the wave equation in the following manner.

Taking only the TT part of the source, $T_{ij}^{TT'}(\eta',\vec{x}')$,
we construct the corresponding part, $\bar{\chi}_{ij}^{TT}(\eta,
\vec{x})$ of the particular solution\footnote{The TT parts is with
	respect to the spatial argument. For the source with argument
$\vec{x}'$ we have used the label $TT'$ while for the $\bar{\chi}_{ij}$
with argument $\vec{x}$ we have used the label $TT$.}. This part
automatically satisfies the spatial TT condition. This is because, the
Green's function is a function of $(\vec{x}-\vec{x}')$ and the partial
derivatives $\partial_x$ can be converted to $-\partial_{x'}$ on the
Green's function. With partial integration these derivatives act on
$T_{ij}^{TT'}(x')$ which satisfies the transversality condition. By
choosing the TT part of the homogeneous solution to be zero, the full
$\chi_{ij}^{TT}$ is obtained.  All the relevant physical (gauge
invariant) quantities can be computed from the $\bar{\chi}_{ij}^{TT}$.
These are the radiative losses, at the linearized level, from the
gravitationally bound source.

One can now do all the usual approximations, $|\vec{x}| \gg |\vec{x}'|$
etc and simplify the TT part of the particular solution. In particular,
in the source integral, keeping only the $o(|x|^{-1})$ terms, 
\begin{eqnarray}
	\bar{\chi}_{ij}^{TT}(\eta, \vec{x}) & \sim & \frac{1}{|\vec{x}|}
	\int d^3x' T^{TT'}_{ij}(\eta - |\vec{x} - \vec{x}'|, \vec{x}')
	\\
	& = & \frac{1}{|\vec{x}|} \int d^3x' \sum_{L\ge
	0}\Big[\partial^{L}_{\eta}T_{ij}^{TT'}(\eta - |\vec{x}|,
	\vec{x}')\Big]\frac{(\hat{x}\cdot \vec{x}')^{L}}{L!} \ .
\end{eqnarray}
The sum is just the Taylor expansion of the stress tensor in the first
argument, $(\eta - |\vec{x} - \vec{x}'|)$ evaluated at $(\eta -
|\vec{x}|)$. Any truncation of the sum, gives a correspondingly
truncated solution and corresponding approximate energy loss. The
leading truncation, in the $\eta$-derivatives, is to take $L=0$. 

The expression for the solution can be put in a more conventional form
using the conservation equation of the stress tensor, which is
guaranteed by the linearized equation as mentioned above. The source
integral is expressed in terms of moments whose definition requires use
of frame components and determinant of the induced metric on the
constant $\eta$ hypersurface. These details are discussed in \cite{DH-1,
ABK-3, CHK}.  Adopting the notation of \cite{CHK} for ease of comparison
we summarize the conservation equations, definitions of moments and
source integrals below.

Using the synchronous time $t$ defined through $\eta := - H^{-1}e^{-Ht}$
so that $\partial_{\eta} = a(t)\partial_t, \ a(t) := e^{Ht}$, the
conservation equations take the form,
\begin{equation} \label{ConservEqn}
	a(t)\partial_t T_{00} = \partial_iT_{0i} - H a(t)\hat{T} ~ ~ , ~
	~ a(t)\partial_tT_{0i} = \partial_jT_{ji} - 2H a(t) T_{0i} \ .
\end{equation}
Let $x_L := x_{i_1}\cdots x_{i_L}$ for $L \ge 0$. Define the {\em
moments},
\begin{eqnarray}
	Q_L(\eta) & := & \int_{\eta=Const}d^3x
	a^{L+1}(\eta)[T_{00}(\eta,\vec{x})] x_L
	\mbox{\hspace{0.9cm}(Mass moments)}\\
	\bar{Q}_L(\eta) & := & \int_{\eta=Const}d^3x
	a^{L+1}(\eta)[\delta^{ij}T_{ij}(\eta,\vec{x})] x_L
	\mbox{\hspace{0.5cm}(Pressure moments)}\\
	P_{i|L}(\eta) & := & \int_{\eta=Const}d^3x
	a^{L+1}(\eta)[T_{0i}(\eta,\vec{x})] x_L
	\mbox{\hspace{1.0cm}(Current moments)}\\
	S_{ij|L}(\eta) & := & \int_{\eta=Const}d^3x
	a^{L+1}(\eta)[T_{ij}(\eta,\vec{x})] x_L
	\mbox{\hspace{1.0cm}(Stress moments)}.
\end{eqnarray}
All moments are automatically symmetric in the $i_1, \cdots, i_L$
indices. We have used $Q_L := Q^{(\rho)}$ and $\bar{Q}_L := Q_L^{(p)}$
for convenience. Note that the pressure moments determine the trace part
of the stress moments.

In terms of these moments, the conservation equations take the form,
\begin{equation}\label{MomentConservEqn}
	\partial_tQ_L = H(LQ_L - \bar{Q}_L) - L P_{(i_1|i_2\ldots i_L)}
	~ ~ , ~ ~ \partial_tP_{i|L} = (L-1)H P_{i|L} - L
	S_{i(i_1|i_2\ldots i_L)} .
\end{equation}
The parentheses indicate symmetrization in the indices. 

Taking $L = 2$ in the first equations and $L=1$ in the second equation
gives,
\begin{eqnarray}
	\partial_tQ_{i_1i_2} & = & H(2 Q_{i_1i_2} - \bar{Q}_{i_1i_2}) -
	2P_{(i_1|i_2)} \ , \\
	\partial_t P_{i|i_1} & = & - S_{ii_1} \ . ~
	\mbox{\hspace{4.0cm}This implies,} \\
	\partial_t^2 Q_{ij} & = & H(2\partial_tQ_{ij} -
	\partial_t\bar{Q}_{ij}) - 2\partial_tP_{(i|j)} \nonumber \\
	\therefore S_{ij}(\eta) & = & \frac{1}{2}\big[\partial_t^2Q_{ij}
	- 2H\partial_tQ_{ij} + H \partial_t\bar{Q}_{ij}\big] \ .
	\label{QPoleEqn}
\end{eqnarray}
A couple of remarks are in order.

{\em Remark-1:} It is more transparent to write the conservation
equation (\ref{MomentConservEqn}) in the form,
\begin{equation}\label{MomnConsEqn}
	H\bar{Q}_L = H LQ_L - \partial_tQ_L - LP_{(i_1|i_2\ldots i_L)} ~
	, ~ L S_{i(i_1|i_2\ldots i_L)} = H(L-1)P_{i|L} - \partial_t
	P_{i|L} ~,~ L \ge 0\ .
\end{equation}
This shows that if we are given {\em mass moments} $Q_L(t)$ and the {\em
current moments} $P_{i|L}(t)$, then the {\em pressure moments}
$\bar{Q}_L(t)$ and the {\em partially symmetrized part} of the stress
moments $S_{ij|L}(t)$ are fully determined by the source conservation
equation (\ref{ConservEqn}).

{\em Remark-2:} The integrands $T_{\mu\nu}x_L$ are tensors on the
constant-$\eta$ hypersurfaces and can be decomposed in terms of their
divergence-free and trace-free parts.  The {\em integrals of these
	decompositions are defined to be the corresponding
decompositions of the moments.} For example, $S_{ij}^{TT}(\eta) :=
\int_{\eta} d^3x a(\eta)T_{ij}^{TT}$ and similarly for others. Since the
definition of moments and the conservation equations are all {\em
linear}, the conservation equations hold for each part separately. Thus
the $TT$ labels can be applied consistently.

In the (\ref{QPoleEqn}), we can put the label $TT$ on both sides.  The
explicit, particular solution $\bar{\chi}_{ij}^{TT}$ coming from the
$L=0$ term in the Taylor expansion of $T_{ij}^{TT}$ then leads to the
quadrupole solution as given in \cite{ABK-3, DH-1}. I repeat that the
other components of $\chi_{\mu\nu}$ are not needed in the computation of
any gauge invariant quantity. The name ``quadrupolar'' comes from the
appearance of the mass and pressure second moments. These are {\em not}
second moments of the $T^{TT}_{ij}$.  The above expressions go over to
the $\Lambda = 0 \ (H \to 0)$ case using the cosmological chart.

The higher order terms in the Taylor expansion of $T_{ij}$, involve
moments of {\em $\eta$-derivatives of $T_{ij}$}. These are {\em not} the
$\eta$-derivatives of the moments, thanks to the explicit factors of
$a^{L+1}(\eta)$ e.g.. the second moments of $\partial_{\eta}^2{T}_{ij}$
are {\em not} the $\partial_{\eta}^2 S_{ij|kl}$. Dropping
$\partial_{\eta}^2{T}_{ij}x^kx^l$ is not the same as dropping
$S_{ij|kl}$. The higher order terms in the Taylor expansion together
with their corresponding moments are easily worked out. For example, in
terms of the $t-$derivatives, 
\begin{equation}
	\int d^3x'\partial^2_{\eta}T_{ij}\ x'_{k}x'_l ~ = ~
	\frac{1}{a}\big[ \ddot{S}_{ij|kl} - 5H \dot{S}_{ij|kl} + 6H^2
	S_{ij|kl}\big] \ .
\end{equation}

Suffice it to say that dropping or keeping these moments {\em does not
lead to any inconsistency with the conservation
equation.}\footnote{However, {\em if} $S_{ij|kl}(t)$ together with its
	$t-$derivatives are taken as negligible, {\em then} the
conservation equations would imply that $\bar{Q}_{ij}$ should also be
neglected\cite{CHK}.\label{Inconsist-footnote}} Since the stress moments
are determined by the mass and current moments, which stress moments can
be neglected is related to the $t-$dependence of the mass and the
current moments. Secondly, though not explicitly stated in the earlier
works, ignoring $\bar{\chi}_{0\mu}$ components of the particular
solution does {\em not} mean that they have been neglected in an
approximation.  They are simply not needed in computing gauge invariant
quantities when complete gauge fixing is employed as explained above. 

To articulate the truncation issue flagged in \cite{CHK}, let us now
change track and take a look at the relation of solutions expressed in
the conformal chart and its transform to the Bondi-Sachs chart.
\section{Transformation to Bondi-Sachs form} \label{BSChrt-sec}
Bondi-Sachs approach is to choose coordinates which capture the idea of
radiation escaping from a bounded region (``isolated body'') to far
away. The radiation is presumed to travel along null geodesics and their
spatial directions could be labeled by points on a sphere surrounding
the isolated body. The points along the geodesics themselves could be
labeled in relation to luminosity distance. This suggests coordinates
$u,r,x^A, A = 1,2$. This presumed interpretation is captured by the
conditions: $g^{uu} = 0$ ($u=$ constant hypersurfaces are null), $g^{uA}
= 0$ (angular coordinates are constant along rays) and
$\partial_r\Case{\mbox{det}(g_{AB})}{r^4} = 0$ ($r$ is the luminosity
distance).  These imply that the covariant components of the metric
satisfy $g_{rr} = 0 = g_{rA}$ \cite{CHK}.  These 3 conditions leave 7
free functions and the metric is conventionally taken in the form, now
on referred to as the BS form,
\begin{equation} \label{BSFormEqn}
	ds^2_{BS} = \Case{V}{r}e^{2\beta}du^2 - 2e^{2\beta}dudr +
	g_{AB}(dx^A - U^Adu)(dx^B - U^Bdu) ~, ~ A, B = 1,2 .
\end{equation}
The functions $V,\beta,g_{AB}, U^A$ are all functions of all the
coordinates: $u \in \mathbb{R}, r > 0$ with $x^A$ being coordinates on
the sphere and may be taken to be the standard $\theta, \phi$. The
radiation is supposed to be detected as $r \to \infty$. The radiative
fields satisfy the source free Einstein equations with a cosmological
constant $\Lambda$. While these choices are adapted to the potential
radiative solutions, not every solution expressed in this form
represents radiation. For instance, the exact de Sitter solution for
$\Lambda > 0$, is non-radiative. This solution is identified in the BS
form by the choices,
\begin{eqnarray}\label{dSinBSEqn}
	\mbox{Choose} & : & \beta = 0 ~,~ \frac{V}{r} = -(1 -H^2r^2) ~,~
	U^A = 0 ~,~ g_{\theta\theta} = 1 ~,~ g_{\phi\phi} = sin^2\theta
	; \\
	\overline{ds}^2_{de Sitter} & = & -(1-H^2r^2)du^2 - 2du dr +
	r^2(d\theta^2 + sin^2\theta d\phi^2) ~ \mbox{(BS coordinates)} .
	\\
	& = & \frac{1}{H^2\eta^2}\big(-d\eta^2 + d\rho^2 +
	\rho^2(d\theta^2 + sin^2\theta d\phi^2)\big) ~ \mbox{(Conformal
	coordinates)}.
\end{eqnarray}
For the de Sitter metric, the conformal chart and the BS chart, are
related as \cite{CHK},
\begin{eqnarray} \label{BSTransformEqn}
	\eta(u,r) = -\frac{e^{-Hu}}{H(1+Hr)} & , &  \rho(u,r) = \frac{r
	e^{-Hu}}{1+Hr} ~ ~ \leftrightarrow \\
	u(\eta,\rho) = -H^{-1}ln\big(H\rho - Hu\big) & , & r(\eta,\rho)
	= -\frac{\rho}{H\eta} \ .
\end{eqnarray}
The radiative solutions are sought within the class of solutions having
certain stipulated asymptotic behaviour as $r \to \infty$. 

A precise stipulation may be achieved with a conformal compactification
which brings the null infinity -- the place where all null geodesics
end-up -- as a boundary, ${\cal I}$. The null infinity is a null
hypersurface for $\Lambda = 0$, is space-like for $\Lambda > 0$ and
time-like for $\Lambda < 0$.  A specific choice of the form of
asymptotic solutions corresponds to a specific definition of {\em
(locally) asymptotically de Sitter} space-times\footnote{`locally'
indicates that the topology of ${\cal I}$ is $\mathbb{R}^3$.  For
asymptotically de Sitter it is $S^3$.}. Unlike the $\Lambda = 0$ case,
the issue of positivity of energy vis a vis characterization of absence
of incoming radiation has not been settled and consequently, there is
some ambiguity left in the choice of boundary conditions i.e. behaviour
as $r \to \infty$.  For definiteness, we will use the stipulation given
in \cite{BBP}. This is obtained by expanding the metric in powers of
$r^{-1}$, demanding reasonable boundary conditions and imposing the
Einstein equations order-by-order.  The result is \cite{CHK, BBP},
\begin{eqnarray} \label{BSAsymFormEqn}
	g_{AB} & = & r^2\gamma_{AB} + r C_{AB} + \frac{C^{CD} C_{CD}
	\gamma_{AB}}{4} + \frac{E_{AB}}{r} + \cdots ~ , ~  \mbox{with}
	\\
	& & \gamma^{AB}C_{AB} = 0 = \gamma^{AB}E_{AB}\ , \
	\mbox{det}(g_{AB}) = sin^2\theta\ ;\\
	\beta & = & \beta_{0} - \frac{1}{32r^2}C^{AB} C_{AB} +
	0(r^{-4}); \\
	U^A & = & U^A_0 - \frac{1}{2r^2}D_BC^{AB} - \frac{2}{3r^3}\Big(
	N^A - \frac{1}{2}C^{AB}D^CC_{BC}\Big) ; \\
	V & = & \frac{\Lambda}{3}r^3 - D_AU^A_0 r^2 - \Big( 1 +
	\frac{\Lambda}{16}C^{AB}C_{AB}\Big) r + 2M + \cdots .
\end{eqnarray}
In the above, the coefficient tensors are functions of $(u, x^A)$.

The two references \cite{CHK,BBP}, differ in the boundary conditions
imposed as a further gauge choice.  Bonga et al choose the two
dimensional metric $\gamma_{AB}$ to be the standard round metric on
sphere and leave $U^A_0$ to be free  while Comp\`ere et al impose only
the determinant condition on the boundary metric $\gamma_{AB}$ and set
$U^A_0 = 0$. For our argument, these differences do not matter (see also
\cite{HHKV}). We will refer to these additional conditions at ${\cal I}$
as the {\em boundary gauge conditions}. Thus, we have the Bondi-Sachs
coordinates together with the stipulated fall off conditions as $r \to
\infty$ specifying locally, asymptotically de Sitter space-times. The
energy, momentum, angular momentum and their fluxes are given in terms
of integrals over $S^2$ at a given $u$ on the null infinity ${\cal I}$
\cite{BBP, CHK}. By transforming the linearized solution obtained in the
conformal chart to the BS chart, we can compute the charges and fluxes
at ${\cal I}$.  This re-expressing of the solution in BS chart is a two
step procedure: first apply the coordinate transformation given in
(\ref{BSTransformEqn}) to bring the background metric in the BS form
followed by an infinitesimal gauge transformation to bring the perturbed
metric in the BS form (\ref{BSFormEqn}). This is described below
\cite{CHK}.

Let us denote the conformal coordinate collectively by $\{x^{\mu}\}$ and
the BS coordinates by $\{\bar{x}^{\mu}\}$.  In the first step, the
conformal chart perturbation $h_{\mu\nu}(x)$ transforms as,
\begin{equation}
\bar{h}_{\mu\nu}(\bar{x}) := \Case{\partial x^{\alpha}}{\partial
\bar{x}^{\mu}} \Case{\partial x^{\beta}}{\partial \bar{x}^{\nu}}
h_{\alpha\beta}(x(\bar{x})) ~ = ~ 
\Case{\partial\eta}{\partial\bar{x}^{\mu}}
\Case{\partial\eta}{\partial\bar{x}^{\nu}} h_{00}+
\Case{\partial\eta}{\partial\bar{x}^{\mu}} \Case{\partial
x^i}{\partial\bar{x}^{\nu}} h_{0i}+
\Case{\partial\eta}{\partial\bar{x}^{\nu}} \Case{\partial
x^i}{\partial\bar{x}^{\mu}} h_{0i}+ \Case{\partial
x^i}{\partial\bar{x}^{\nu}} \Case{\partial x^j}{\partial\bar{x}^{\mu}}
h_{ij}
\end{equation}
One can see that $\bar{h}_{rr} \neq 0$ in general. It is non-zero even
if we take $h_{0\beta} = 0$ (synchronous gauge) and take only the TT
part of the $h_{ij}$.  Hence, this transformed metric
$(\bar{g}_{\mu\nu})_{BS} + \epsilon\bar{h}_{\mu\nu}$ does not conform to
the BS form and a further coordinate transformation is needed to bring
it to the BS form.  This extra transformation must be an infinitesimal
gauge transformation since only the perturbation violates the BS form.
It is thus convenient not to do complete gauge fixing in the conformal
chart prior to the second transformation. The components $h_{0\mu}$ are
then non-zero.  Up to this point, no $log(r)$ terms are generated.  It
should be remembered that the $h_{\mu\nu}(x)$ to which coordinate
transformations are to be applied is the sum
$\Case{1}{H^2\eta^2}(\bar{\chi}_{\mu\nu} + \underline{\chi}_{\mu\nu})$
which satisfies the generalized harmonic gauge condition. 

Let $X^{\mu}(\bar{x}) := \bar{x}^{\mu} + \epsilon\zeta^{\mu}(\bar{x})$
denote the new coordinates and let $H_{\mu\nu}(X)$ denote the gauge
transformed perturbation: $H_{\mu\nu}(X) := \bar{h}_{\mu\nu}(X) +
\epsilon{\cal L}_{\zeta}\bar{g}_{\mu\nu}(X)$. The $\zeta^{\mu}$ is
chosen such that the BS coordinate conditions (A): $H_{rr} = H_{rA} =
\partial_r(\gamma^{AB}H_{AB}) = 0$ are satisfied along with the
supplementary boundary gauge conditions (B): $r^{-2}\gamma^{AB}H_{AB}
\to 0, \ r^{-2}H_{uu} \to 0, \ r^{-2} H_{uA} \to 0$ as $r \to \infty$.  

The BS conditions (A) give first order, inhomogeneous differential
equations for $\zeta^{\mu}$. The corresponding particular solutions are
determined in terms of the transformed $\bar{h}_{\mu\nu}(X)$ and the
(undetermined) homogeneous solutions, $\xi^{\mu}$.  Next, the
non-vanishing components of $H_{\mu\nu}$ must satisfy the asymptotic
forms of general asymptotically de Sitter solution (\ref{BSAsymFormEqn})
together with the boundary functions. These give first order,
inhomogeneous differential equations for $\xi^{\mu}$ \cite{CHK}.  The
homogeneous part of the solutions $\xi^{\mu}$ give the $\Lambda$-BMS
generators while the particular solutions involving the transformed
$\bar{h}_{\mu\nu}(X)$, yield the boundary functions in the limit $r \to
\infty$. From these arguments, it should be clear that the asymptotic
form of $\bar{g}_{\mu\nu}(X) + \epsilon H_{\mu\nu}(X)$ is directly
controlled by the conformal chart solution $\bar{h}_{\mu\nu}(\bar{x})$
that is fed into the transformation procedure.

In \cite{CHK}, the conformal chart solution was taken to be the {\em
particular solution} $(H\eta)^{-2}\bar{\chi}_{\mu\nu}$. It was observed
that if one uses the truncation, $S_{ij|kl} = 0$ and does not drop the
$\bar{Q}_{ij}$ terms from the particular solution (see the footnote
\ref{Inconsist-footnote}), then the transformed solution contains
$log(r)$ terms.  The occurrence of $log(r)$ is an inconsistency since
for $\Lambda > 0$ Einstein equations forbid their occurrence\footnote{I
	thank Amitabh Virmani for pointing this out.  Incidentally, in
	the $\Lambda=0$ case, $log(r)$ terms are permitted by Einstein
	equations modulo the choice of smoothness stipulations on the
unphysical metric $\Omega^2 g$ near the null infinity \cite{GWGR14}.}.
The same procedure applied to the $\Lambda=0$ case generates no such
terms.  It was further checked \cite{CHK} that if the quadrupolar
truncation, defined as keeping up to $L=2$ terms in Taylor expansions of
all components of $T_{\mu\nu}$, is used, then the log terms disappear.
This led the authors of \cite{CHK} to flag the issue of {\em consistency
of truncation.} 

The conformal chart computation done with the fully gauge fixed solution
hints at no obstruction to any truncation. So why does an issue arise
when transformation to BS form is done?  Notice that the occurrence of
the $log(r)$ terms in the final transformed perturbation could also be
an indication that {\em the perturbation $\bar{h}_{\mu\nu}(X)$ that we
feed in, is {\em not} the solution of the linearized equation at all!}
Recall that $\chi_{\mu\nu}$ is a solution {\em provided} it satisfies
the wave equation {\em and} the generalized harmonic gauge condition.
Since the particular solution satisfies the wave equation, it may fail
to satisfy the generalized gauge condition. For convenience, here is the
particular solution \cite{VRS, CHK}, 
\begin{eqnarray}\label{ParticularSoln}
	\hat{\chi}(\eta,\vec{x}) & = & 4\int_{source}\frac{d^3x'}{R}
	\frac{\eta}{\eta-R} \hat{T}(\eta-R,\vec{x}') ~,~ R :=
	|\vec{x}-\vec{x}'| ; \\
	\chi_{0i}(\eta,\vec{x}) & = & 4\int_{source}\frac{d^3x'}{R}
	\frac{\eta}{\eta-R} T_{0i}(\eta-R,\vec{x}') ~  ; \\
	\chi_{ij}(\eta,\vec{x}) & = & 4\int_{source}\frac{d^3x'}{R}
	T_{ij}(\eta-R,\vec{x}') ~+~ 4\int_{source}d^3x'
	\int_{-\infty}^{\eta-R} \frac{d\eta'}{\eta'}\partial_{\eta'}
	T_{ij}(\eta',\vec{x}')   . 
\end{eqnarray}
The generalized gauge condition is given in (\ref{GaugeCondn}).

It is easy to check that the particular solution given above satisfies
only the $\mu = 0$ part of the gauge condition and {\em fails to satisfy
the $\mu=i$ part of the gauge condition}.  Hence the particular solution
above is not a solution of the linearized Einstein equation by itself.
Incidentally, for the $\Lambda=0$ case, the generalized harmonic gauge
condition reduces to the harmonic gauge condition and the corresponding
particular solution does satisfy it automatically. That particular
solution {\em is} a solution of the linearized equation and we don't
encounter any $log(r)$ terms emerging. 

Clearly, the way out, for positive $\Lambda$, is that suitable
homogeneous solution $\underline{\chi}_{\mu\nu}$ should be added so as
to satisfy the generalized harmonic gauge condition. If the particular
solution is truncated, the corresponding homogeneous solution would also
be truncated accordingly.  The so chosen total solution should be taken
for the transformation to the BS form and no $log(r)$ terms should
appear for any truncation.  Equally well, if we choose the fully gauge
fixed solution $\chi_{ij}^{TT}, \chi_{0\nu} = 0$ and followed the
transformation procedure, no $log(r)$ terms should appear. However,
explicit check of these statements is not undertaken in this brief note.

To summarize: The generalized harmonic gauge choice points out the
possibility that particular solution of the wave equation may not
satisfy the gauge condition imposed in arriving at the wave equation and
care is needed to ensure that one is dealing with a solution of the
linearized equation. Solutions of the homogeneous wave equation are
available for ensuring this. As a precursor to a ``post-de Sitter''
formalism, it is useful to the transform the solution from the conformal
chart to the Bondi-Sachs chart as done in \cite{CHK}, since the physical
charges and fluxes are well specified in this chart.  Conformity of the
transformed solution to the Bondi-Sachs form is a useful consistency
check on the computations.

I end with a few general remarks.

The present value of cosmological constant, in geometrized units, is
about $\sqrt{\Lambda} \sim 10^{-26}$ (meter)$^{-1}$. For isolated
sources, we have a length scale related to its mass/size of the
corresponding Schwarzschild radius, $R_S$. Even for a trillion solar
mass black hole, this corresponds to about $10^{15} m$. The wave
length/frequency is another length scale, which even for nano-Hertz
waves (astrophysical gravitational background) is only  about $\lambda
\sim 10^{17} m$. The distance between the source and the detector (us),
at giga-parsec is about $r \sim 10^{24} m$. Among the corresponding
dimensionless numbers (corrections due to positive $\Lambda$) are,
respectively, $10^{-11}, 10^{-9}, 10^{-2}$. It should be noted however
that the corrections must have positive powers of $\Lambda$ since we
have not seen any drastic effects of possible presence of non-zero
cosmological constant in the detected gravitational waveforms. The
distance $r$ however, has to be in the denominator, since we are dealing
with radiation from isolated source. Hence the dimensionless corrections
must use only the source scale and wavelength scales and these are very
small, $10^{-9}$ or smaller at the level of wave amplitude.

If one wants to consider a ``post-de Sitter'' formalism analogous to the
post-Minkowski formalism, then at least the usual PN series of near
source solutions should be valid to a very good approximation. The
asymptotic form of the solutions however has a clear imprint of the
cosmological constant. These two solutions are to be appropriately
matched in an intermediate region. It is from the equations of motion
with energy loss included, that the cosmological constant would show up
even if $\Lambda$ is neglected in the local conservative dynamics. Be as
it may, the argument based on the scales still implies that these
corrections will be very small in the gravitational radiation from
isolated sources.  Cosmological gravitational waves (the
astrophysical/primordial gravitational wave background) is a different
ball game.

I would like to thank the organizers of the conference for the
invitation which gave me the opportunity for this clarificatory note. I
also thank Jahanur Hoque and Amitabh Virmani for discussions and
comments on the draft of this note.

\end{document}